\documentclass[%
 reprint,
 amsmath,amssymb,
 aps,
 pra,
]{revtex4-2}

\usepackage[english]{babel}
\usepackage[utf8]{inputenc}
\usepackage[T1]{fontenc}
\usepackage{amsmath}
\usepackage{amssymb}
\usepackage{mathtools}
\usepackage{graphicx}
\usepackage{dcolumn}
\usepackage{bm}
\usepackage{xcolor}
\usepackage[labelformat=simple]{caption}
\newcommand{\ee}{\mathrm{e}}
\newcommand{\dd}{\mathrm{d}}
\newcommand{\out}{\mathrm{out}}
\newcommand{\inc}{\mathrm{in}}
\newcommand{\refl}{\mathrm{ref}}
\DeclareMathOperator{\sgn}{sgn}

\begin{document}

\title{A Passive Intelligent Metasurface: \\ A Network Model}

\author{Tristan Lawrie}
\affiliation{Centre for Metamaterial Research and Innovation, Department of Physics and Astronomy, University of Exeter, Exeter EX4 4QL, United Kingdom}
\author{Gregor Tanner}
\affiliation{School of Mathematical Sciences, University of Nottingham, Nottingham NG7 2RD, United Kingdom}
\author{Luke G. Bennetts}
\affiliation{School of Mathematics and Statistics, The University of Melbourne, Victoria 3010, Australia}

\date{\today}

\begin{abstract}
Intelligent metasurfaces enable advanced wave manipulation, including negative refraction, focusing, beam splitting and beam steering, with significant potential for practical applications. However, conventional approaches often rely on active metamaterials with gain, or demand computationally intensive solutions of inverse problems to find the optimal configuration. In this work, we use the scattering framework of quantum graph theory to derive an exact, analytic expression for configuration of a passive metasurface that produces a prescribed reflected wave field. The metasurface terminates a periodic lattice of waveguides and consists of an array of channels, each a tube of length $\ell_m$ ending in a vertex whose scattering parameter $\lambda_m$ sets how much of the wave is lost to the wall. For any prescribed incident and reflected field, the exact lengths and scattering parameters follow channel by channel, in closed form, from the ratio of the wave amplitudes returning to and leaving the lattice, and the metasurface is passive whenever this ratio does not exceed unity in modulus. We demonstrate the construction for both a one-dimensional and a two-dimensional metasurface. The one-dimensional metasurface terminates a two-dimensional lattice and steers, splits or perfectly absorbs Gaussian beams. The two-dimensional metasurface terminates a three-dimensional lattice and generates a prescribed image in a plane inside the lattice. In each case, we obtain the exact boundary properties that produce the effect.
\end{abstract}

\maketitle

\section{Introduction}\label{sec:intro}

Wave manipulation and wave control has emerged as an important research field, with applications in optics, wireless communication and acoustics, among other fields~\cite{rotter2017,Kildishev2013,Assouar2018,Basar2019}.
The core question is, given a source and a (possibly complex or disordered) medium, how much control can be exerted over the resulting wave field by acting on a finite, structured set of degrees of freedom? Over the past two decades this question has been answered using linear wave theory and ideas from metamaterial design based on the following principles: a linear wave operator; a scattering or transfer matrix connecting inputs to outputs; and an optimisation problem over a constrained control space.

The seminal demonstrations of wavefront shaping through and around complex scattering media showed that even strongly multiple-scattering systems --- normally scrambling coherent information irrecoverably --- admit a well-defined input--output linear map, known as the transmission matrix, whose singular value decomposition reveals high-transmission ``open channels''~\citep{vellekoop2007, vellekoop2008, popoff2010,MoskLeroseyFink2012NatPhoton}. Shaping the incident wavefront via spatial light modulators (SLM)~\cite{Leonardo2007, Geng2013}, spatial microwave modulators (SMM)~\cite{Zhou2018} or an acoustic analogon~\cite{Memoli17, Memoli20}, allows one to send wave energy through opaque scattering layers thereby synthesising prescribed output patterns~\citep{kim2012, rotter2017}. This body of work, originally developed for optical speckle, has since been transposed to the microwave~\cite{Ambichl2017} and acoustic regimes~\cite{Rivet2018}, where the number of controllable degrees of freedom is more modest.

In a reverberant setting, such as wave fields in a room or cavity, multiple scattering typically leads to randomness of the wave field~\cite{beenakker1997}.
Here a different approach is tried and tested. Instead of shaping the incoming wave field to suit a given environment, the environment is itself transformed to provide a specified cavity wave field for a given source distribution~\citep{kaina2014, Dupre2015, Gross2020, Anlage2020}. It has been shown to lead to controlled enhancement of the diversity and capacity of the communication channels both for microwave~\citep{delhougne2019} and acoustic~\citep{Ma2018, Zhang2024} communication. The environment shaping effect is typically achieved by installing subwavelength-structured interfaces or panels --- metasurfaces --- capable of imposing an essentially arbitrary local phase, amplitude, or polarization response on an incident wave. Reconfigurability is achieved by embedding active elements (varactors, PIN diodes, MEMS switches, or their acoustic analogues) into each meta-atom, which in the simplest and most robust implementation yields a \emph{binary} response, in which each unit cell can only be switched between two discrete states~\citep{zhang2021}. This restriction, while attractive from a fabrication and control standpoint, converts the continuous wavefront-shaping optimization into a combinatorial one.

An engineering counterpart of this program has developed, in the electromagnetic domain, around reconfigurable intelligent surfaces (RIS): large, low-cost, nearly-passive arrays of tuneable scattering elements deployed to shape the radio propagation environment for wireless communication, effectively turning the wireless channel itself into an optimisation variable~\citep{DiRenzo2019,wu2020smart, direnzo2020, alexandropoulos2020}. Practical RIS hardware is almost universally quantised to a small number of phase states, and the resulting discrete beamforming problem inherits the same combinatorial structure as binary coding metasurfaces~\citep{wu2020smart, zhang2021}. The acoustic analogue --- reconfigurable acoustic metasurfaces used to control sound fields in rooms, ducts, or underwater channels --- has been demonstrated to provide multi-channel, cross-talk-free acoustic communication in reverberant rooms using binary-controlled acoustic reconfigurable metasurfaces~\citep{Zhang2024}.

The wave shaping scenarios mentioned above have one thing in common: In order to achieve the desired wave forms, optimisation algorithms have to be employed to find the best metasurface configuration for the task at hand. In this paper, we will use a description of tuneable metasurfaces in terms of \emph{quantum graphs}. We will demonstrate that an exact map between incoming and outgoing reflected and transmitted waves can be obtained in this setting. Further, we will show that the process is a solvable inverse problem, that is, we can uniquely determine the parameters of the metasurface knowing the incoming and desired outgoing field, thus sidestepping an optimisation algorithm.

A quantum graph is a network of one-dimensional edges (waveguides, pipes, or transmission lines) meeting at vertices; free one-dimensional wave propagation along edges is considered with graph vertices acting as low-dimensional, unitary scattering centres~\citep{kottos1999,Gnutzmann2006,berkolaiko2013}. Quantum graphs have served in mesoscopic and wave-chaotic physics as a computationally cheap model for wave-scattering dynamics. The response of a quantum graph, including its Green's function for a forcing, can be assembled through relatively small matrix problems rather than through full-wave simulations~\citep{lawrie2023}. Building on this correspondence, Lawrie et al.~\citep{lawrie2022} showed that a periodic quantum graph, with finite sub-graphs attached at each vertex, mimicking subwavelength resonant elements, reproduces the band structure and exotic dispersion features of genuine electromagnetic metamaterials while remaining tractable analytically. The same framework was subsequently extended to acoustic metasurfaces built from networks of interconnected pipes, where tuning the graph topology and edge lengths was shown to engineer non-resonant negative refraction of surface waves, confirmed both in full-wave simulations and in experiments on acoustic pipe networks~\citep{lawrie2024}. Choosing the edge lengths alone can even engineer a Doppler effect in a static material, producing vortices and black-, white- and worm-hole-like wave trajectories~\cite{lawrie2026doppler}. Modelling interfaces with graph methods has shown that structures placed between interfaces can be used to engineer their reflection and transmission coefficients~\cite{lawrie2023interface}, and has led to the discovery of angular filtering, which has been validated by numerical simulations~\cite{lawrie2025} and by microwave experiments on networks of coaxial cables, which showed unit transmission at a designed angle~\cite{lawrie2026}. Extending the formulation to a gauge-shifted wave equation, the magnetic Schr\"odinger equation of a charged particle, makes the transmission angle tunable by a magnetic flux~\cite{lawrie2025flux}, and the transmission profile of a waveguide network can be prescribed exactly, with the network constructed analytically~\cite{Lawrie2026transmission}. A detailed account of the formalism behind these works is given in Ref.~\cite{lawriethesis}. In the present work we adopt this quantum graph picture as the underlying model for our tunable metasurface interface which acts as a reflecting \emph{wall}: each meta-atom is represented as a finite graph whose vertex-scattering data can be tuned a priori to provide the desired transmission matrix.

Physically, the wall we have in mind is an array of resonant channels, each a tube of variable length terminated by a tunable scatterer, see Fig.~\ref{fig: Reflection Schematic}. It is close to the experimental metasurfaces of acoustic cavity resonators that have been built to dissipate electric-motor noise in several frequency bands~\cite{tian2022motor}: the tubes of our wall play the role of those cavities, and their lengths and terminations are the quantities we design. Arrays of resonators with different, graded geometries are known to produce rich scattering responses~\cite{Bennetts2018,meng2021rainbow}; here the grading is chosen exactly to produce a prescribed field.


The paper is organised as follows. Section~\ref{sec:formulation} introduces the model and Sec.~\ref{sec:vertex} inverts the boundary condition of a single channel. Section~\ref{sec:lattice} describes the Bloch waves of the lattice and the scattering orders of a periodic wall, and Sec.~\ref{sec:parameterizing} gives the design rule for an arbitrary prescribed field. We apply it to Gaussian beams that are steered, absorbed and split by the wall in Sec.~\ref{sec:results}, and to image formation by a two-dimensional metasurface in Sec.~\ref{sec:3D}, before concluding in Sec.~\ref{sec:conclusion}. All derivations are collected in the appendices.

\section{Formulation}\label{sec:formulation}

\begin{figure*}[t]
\centering
\includegraphics[width = 1\textwidth]{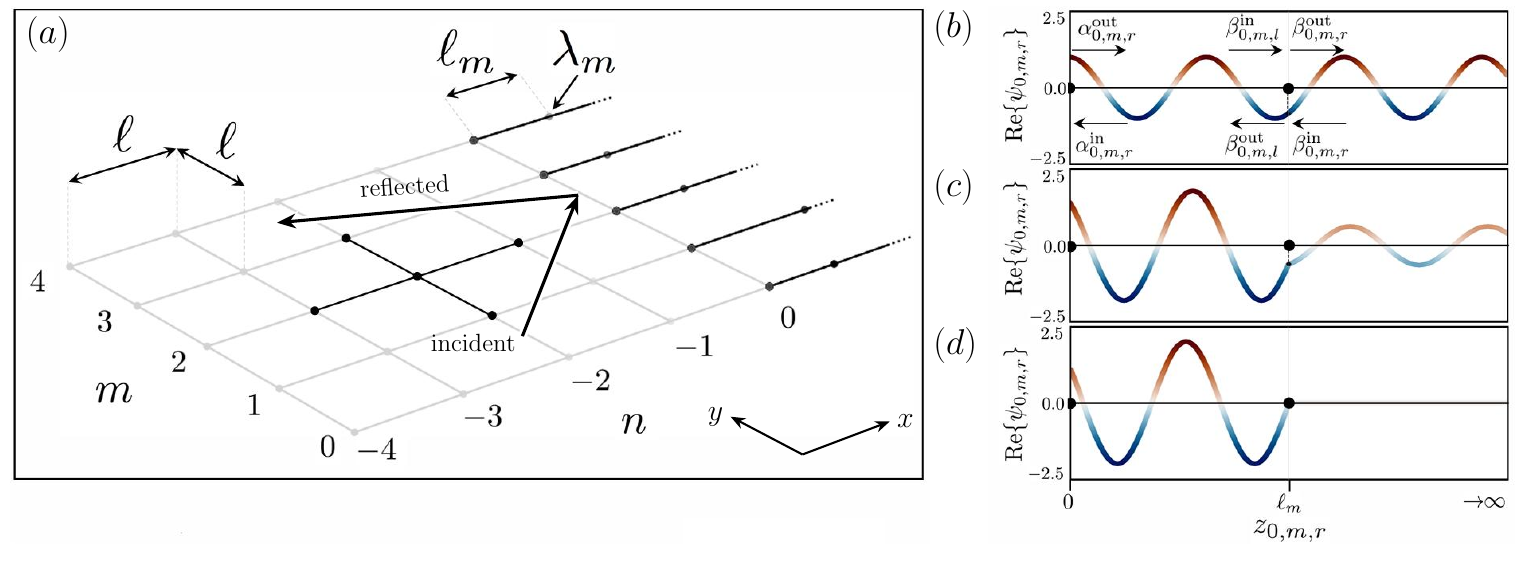}
\caption{(a) A unit-cell strip of the semi-infinite square lattice, with edges of length $\ell$ and coordinates $x = n\ell$, $y = m\ell$, terminated at $n = 0$ by a wall of $M = 5$ channels. Each channel is a tube of length $\ell_m$ ending in a vertex with parameter $\lambda_m$, beyond which a semi-infinite lead (dotted) carries waves away from the lattice; the arrows indicate an incident and a reflected wave. (b)--(d) The field $\mathrm{Re}\{\psi_{0,m,r}\}$ along a single channel, $z_{0,m,r}\in[0,\infty)$, excited with $\alpha^{\text{out}}_{0,m,r} = 1$ and $\beta^{\text{in}}_{0,m,r} = 0$, for (b) $\lambda_m = 0$, (c) an intermediate value of $\lambda_m$ and (d) $\lambda_m\to\infty$; the channel vertex sits at $z_{0,m,r} = \ell_m$. Panel (b) labels the amplitudes travelling towards and away from the wall vertex and the channel vertex.}\label{fig: Reflection Schematic}
\end{figure*}

We consider a semi-infinite square lattice of thin waveguides (edges) of length $\ell$, with vertices labelled $(n,m)$, $n\le 0$, $m\in\mathbb{Z}$, at the positions $x = n\ell$ and $y = m\ell$, see Fig.~\ref{fig: Reflection Schematic}(a). The four edges meeting at a vertex are labelled by the direction in which they point, $e\in\{l,r,d,u\}$ (left, right, down, up); on edge $e$ of vertex $(n,m)$, we introduce a local coordinate $z_{n,m,e}\in[0,\ell]$ measured from the vertex. On every edge, the wave field satisfies the one-dimensional Helmholtz equation
\begin{equation}\label{eq:helmholtz}
\frac{\dd^2\psi_{n,m,e}}{\dd z_{n,m,e}^2} + k^2\psi_{n,m,e} = 0 .
\end{equation}
This equation describes acoustic waves in narrow pipes, transverse waves on strings, voltages and currents on transmission lines and in microwave networks, and electrons in quantum wires~\cite{Kuchment2002,Gnutzmann2006,berkolaiko2013,Hul2004,lawrie2024}. The time dependence is given by the phase $\ee^{-i\omega t}$, where the relation between $\omega$ and $k$ depends on the physical system being modelled.

At every vertex, the edges are coupled by two conditions: (1)~the field is continuous, $\psi_e(0) = \psi_{e'}(0)\equiv\phi$ for any two edges $e$ and $e'$ meeting at the vertex; and (2)~the outgoing derivatives satisfy $\sum_e \dd\psi_e/\dd z_e(0) = \lambda\phi$, with a real vertex parameter $\lambda$~\cite{berkolaiko2013}. Writing the field on each edge as a wave leaving and a wave arriving at the vertex, standard algebra turns these conditions into a vertex scattering matrix that maps the incoming onto the outgoing amplitudes (Appendix~\ref{app:channel} and Ref.~\cite{lawriethesis}), such that
\begin{equation}\label{Scattering Definition}
\boldsymbol{S}(k;\lambda) = \frac{2ik}{vik - \lambda}\mathbf{E} - \mathbf{I},
\end{equation}
where $\mathbf{E}$ is the $v\times v$ matrix of ones, $\mathbf{I}$ the identity and $v$ the valency of the vertex. In this paper, we have $v = 4$ or $v=6$ for a 2D or 3D lattice and we generally assume Neumann--Kirchhoff junctions, that is, $\lambda = 0$. General $\lambda$ values can be interpreted, for example, as transverse waves on a network of strings under tension $T$, a point mass $\mu$ and a grounded spring of stiffness $s$ at the vertex. This then gives $\lambda = (s-\mu\omega^2)/T$, so that negative values of $\lambda$ are mass-like and positive values spring-like.

The lattice is terminated at column $n=0$ by the {wall}. The right-hand edge of every vertex $(0,m)$ is replaced by a tube of length $\ell_m$ that ends in a vertex of valency two with parameter $\lambda_m$, beyond which a semi-infinite lead carries waves away from the structure [Fig.~\ref{fig: Reflection Schematic}(a)]. The lead represents any mechanism that removes energy from the lattice, such as an anechoic termination, an absorbing material or a port into another system; $\lambda_m$ determines how much of the wave is lost to the wall [Fig.~\ref{fig: Reflection Schematic}(b)--(d)]. For $\lambda_m = 0$ the vertex is transparent and the wave passes straight through, while as $\lambda_m\to\infty$ the vertex imposes a Dirichlet condition and the wave is fully reflected. We take the wall to be periodic with period $L = M\ell$. One period thus contains $M$ channels with $(\ell_{m+M},\lambda_{m+M}) = (\ell_m,\lambda_m)$.

\section{Designing the channels}\label{sec:vertex}

Each channel leaving from a vertex $(0,m)$ consists of a tube of length $\ell_m$ terminated by a vertex with $v=2$ which in turn is connected to an infinite lead, see 
Fig.~\ref{fig: Reflection Schematic}(b). Let $\alpha^{\out}_{0,m,r}$ be the amplitude of the wave leaving the wall vertex $(0,m)$ along its right-hand edge $r$, and $\alpha^{\inc}_{0,m,r}$ the amplitude returning to it. For $v = 2$, the scattering matrix~\eqref{Scattering Definition} at the channel vertex has reflection coefficient $\rho_m = \lambda_m/(2ik-\lambda_m)$. A wave entering the tube is reflected at the channel vertex and returns to the wall vertex. We require that the wall is passive, that is, no wave amplitude enters from the lead; we thus find that the two amplitudes $\alpha^{\inc/\out}_{0,m,r}$ are related by (Appendix~\ref{app:channel})
\begin{equation}\label{Governing equation}
\ee^{2ik\ell_m}\rho_m = \frac{\alpha^{\inc}_{0,m,r}}{\alpha^{\out}_{0,m,r}} \equiv \Gamma_m .
\end{equation}
This is the key equation of this work. For any prescribed field in the lattice the amplitudes on each tube, and hence $\Gamma_m$, are known, and Eq.~\eqref{Governing equation} is a single complex condition on the two real parameters $\ell_m$ and $\lambda_m$. Its modulus fixes the vertex parameter and its phase fixes the tube length, which gives (Appendix~\ref{app:channel})
\begin{align}
\lambda_m &= \sigma_m\,\frac{2k\,|\Gamma_m|}{\sqrt{1-|\Gamma_m|^2}}, \label{eq:lambda}\\
\ell_m &= \frac{1}{2k}\Big[\arg\Gamma_m + \sigma_m\arccos\!\big(-|\Gamma_m|\big)\Big] + \frac{q\pi}{k}, \label{eq:length}
\end{align}
with $\sigma_m = \pm 1$ and $q\in\mathbb{Z}$. The vertex parameter is real, and the channel passive, if and only if $|\Gamma_m|\le1$, that is, if the channel absorbs rather than emits energy. A channel that must absorb everything ($\Gamma_m = 0$) is an open lead, $\lambda_m = 0$, whose length is irrelevant, while a channel that must reflect everything ($|\Gamma_m| = 1$) is a Dirichlet channel, $\lambda_m = \pm\infty$, whose phase is set by the length alone. Throughout we take $\sigma_m = -1$, so that every scatterer is mass-like; the integer $q$ allows any minimum tube length to be imposed.

\section{Lattice solutions}\label{sec:lattice}

\begin{figure*}[t]
\centering
\includegraphics[width = 1.0\textwidth]{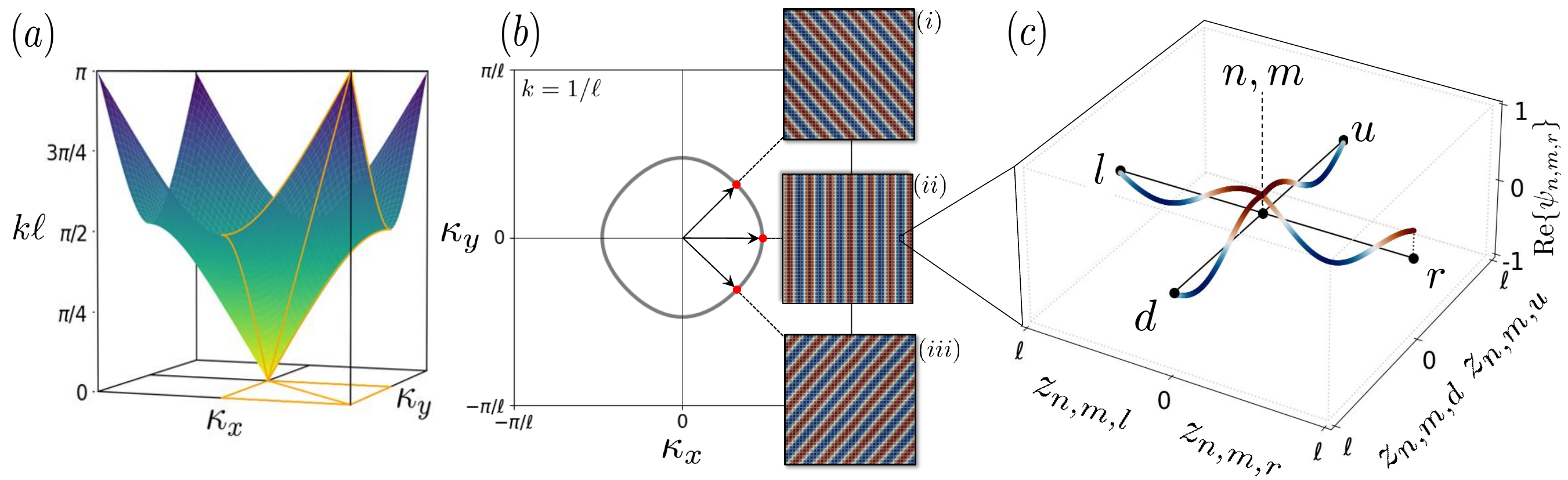}
\caption{(a) Band diagram $k\ell(\kappa_x,\kappa_y)$ of the square periodic lattice, Eq.~\eqref{eq:dispersion2D}. (b) Iso-frequency contour at $k = 1/\ell$, showing the right-travelling solutions for three values of $\kappa_y$: (i) $\kappa_y = 1/\ell$, (ii) $\kappa_y = 0$ and (iii) $\kappa_y = -1/\ell$; the insets show the corresponding Bloch wave fields on the lattice. (c) Neighbourhood of a vertex $(n,m)$ of the lattice in inset (ii), with the directions $e \in \{l,r,d,u\}$ and the edge coordinates $z_{n,m,e}$ marked. The real part of the wave function, $\mathrm{Re}\{\psi_{n,m,e}(z_{n,m,e})\}$, is plotted on each edge.}\label{fig: Lattice Solutions}
\end{figure*}

To express the tube amplitudes through the lattice field we need the Bloch waves of the lattice. The solutions of the square periodic quantum graph are well known~\cite{lawrie2022,lawrie2024,lawrie2025,lawriethesis}, and we only state the key results here; details are given in the references and in Appendix~\ref{app:bloch}. 
Note that the lattice is simply a convenient way of representing the waves that scatter from the wall.
Although the formalism is discrete, its solutions translate directly to the continuum: the discrete angular filter of Ref.~\cite{lawrie2025}, designed on a lattice, agreed excellently with full-wave simulations of the corresponding continuous structure. For this reason we use the discrete formalism throughout.

At every vertex the amplitudes leaving along the edges $(l,r,d,u)$ follow from those arriving through $\mathbf{a}^{\out} = \boldsymbol{S}\mathbf{a}^{\inc}$, with $\boldsymbol{S}$ from Eq.~\eqref{Scattering Definition} for $\lambda = 0$ and $v = 4$. By writing the incoming amplitudes at a vertex in terms of the outgoing amplitudes of its neighbours, $\mathbf{a}^{\inc} = \boldsymbol{P}\mathbf{a}^{\out}$, via the phase matrix $\boldsymbol{P}$, which contains the phase $\ee^{ik\ell}$ accumulated along each edge and the Bloch phases $\ee^{\pm i\kappa_x\ell}$ and $\ee^{\pm i\kappa_y\ell}$ between neighbouring vertices (Appendix~\ref{app:bloch}), we construct the eigenvalue problem
\begin{equation}\label{eq:eig}
\big[\mathbf{I} - \boldsymbol{P}(k,\kappa_x,\kappa_y)\boldsymbol{S}\big]\mathbf{a}^{\inc} = 0.
\end{equation}
Nontrivial solutions exist when $\det[\mathbf{I}-\boldsymbol{P}\boldsymbol{S}] = 0$, which gives the dispersion relation [Fig.~\ref{fig: Lattice Solutions}(a)]
\begin{equation}\label{eq:dispersion2D}
\cos(\kappa_x\ell) + \cos(\kappa_y\ell) = 2\cos(k\ell).
\end{equation}
In the long-wavelength regime, $k\ell\ll1$, expanding the cosines gives $\kappa_x^2 + \kappa_y^2 = 2k^2$, so that up to a factor of two the lattice reproduces the free-space dispersion relation. For a given $k$ and $\kappa_y$, Eq.~\eqref{eq:dispersion2D} gives a pair $\pm\kappa_x$, corresponding to a wave travelling towards the wall and a wave travelling away from it, or to a pair of evanescent waves [Fig.~\ref{fig: Lattice Solutions}(b)]. The eigenvectors of Eq.~\eqref{eq:eig} give the field on every edge [Fig.~\ref{fig: Lattice Solutions}(c)].

A wall of period $L = M\ell$ preserves the Bloch phase of the incident wave modulo $2\pi/L$. An incident wave with tangential wavenumber $\kappa_y^{(0)}$ is, therefore, scattered into the $M$ orders
\begin{equation}\label{eq:orders}
\kappa_y^{(j)} = \kappa_y^{(0)} - \frac{2\pi j}{L}, \qquad j \in \{0,1,\dots,M-1\},
\end{equation}
each of which propagates or decays into the lattice depending on whether Eq.~\eqref{eq:dispersion2D} gives a real or a complex $\kappa_x^{(j)}$. The field in the lattice is the incident Bloch wave, with amplitude $A$, plus a sum over the reflected orders with amplitudes $B^{(j)}$, and the choice of the $B^{(j)}$ defines the reflected field we wish to produce.

\section{Designing the wall}\label{sec:parameterizing}

Evaluating the prescribed field on the tube attached to each wall vertex gives $\alpha^{\out}_{0,m,r}$ and $\alpha^{\inc}_{0,m,r}$, and, therefore, $\Gamma_m$. This is most transparent in terms of the field at the wall vertex, $\Phi_m$, and its continuation $\tilde\Phi_m$ to a fictitious column $n = 1$, the value the missing lattice vertex would have had. Using Eq.~\eqref{eq:orders}, both are discrete Fourier sums over the scattering orders,
\begin{align}
\Phi_m &= \ee^{i\kappa_y^{(0)}m\ell}\Big[A + \sum_{j} B^{(j)}\ee^{-2\pi i jm/M}\Big], \label{eq:Phi}\\
\tilde\Phi_m &= \ee^{i\kappa_y^{(0)}m\ell}\Big[A\,\ee^{i\kappa_x^{(0)}\ell} + \sum_{j} B^{(j)}\ee^{-i\kappa_x^{(j)}\ell}\ee^{-2\pi i jm/M}\Big], \label{eq:Phitilde}
\end{align}
with $j$ running over the $M$ orders~\eqref{eq:orders}. Expressing the tube amplitudes through these sums (Appendix~\ref{app:design}) gives
\begin{equation}\label{eq:Gamma}
\Gamma_m = \frac{\ee^{ik\ell}\Phi_m - \tilde\Phi_m}{\tilde\Phi_m - \ee^{-ik\ell}\Phi_m}.
\end{equation}
Equations~\eqref{eq:lambda}--\eqref{eq:Gamma} form the complete design rule, such that we can choose the reflected amplitudes $B^{(j)}$, evaluate two discrete Fourier transforms, and read off $\lambda_m$ and $\ell_m$ for every channel. Each channel then presents to its wall vertex exactly the response that the missing lattice column would have presented to the prescribed field, so that the prescribed field is the exact response of the designed wall.

\begin{figure}[t]
\centering
\includegraphics[width = 1.0\columnwidth]{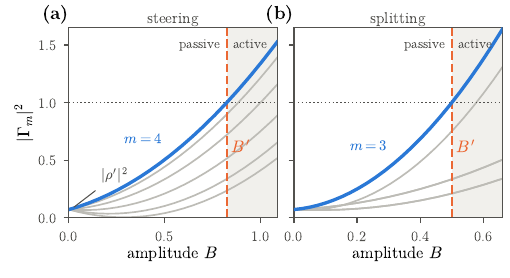}
\caption{$|\Gamma_m|^2$ of every channel as a function of the amplitude $B$ of the reflected field, for (a) the steering and (b) the splitting walls of Fig.~\ref{fig:results}. At $B = 0$ all channels take the perfect-absorber value $|\rho'|^2$. The wall is passive while every channel lies below $|\Gamma_m|^2 = 1$; the first channel to reach the line sets $B'$, beyond which (shaded) the wall would require gain.}\label{fig:passivity}
\end{figure}

The amplitudes $B^{(j)}$ are free parameters. Writing $B^{(j)} = B\hat B^{(j)}$ for a chosen profile $\hat B^{(j)}$, the overall amplitude $B$ sets how much energy is reflected. By tuning $B$ down we ensure that the reflected field carries less energy than the incident field and that no channel has to emit energy, which results in a purely passive reflecting boundary (Fig.~\ref{fig:passivity}). The largest such amplitude, $B'$, is reached when the first channel reflects perfectly, $|\Gamma_m| = 1$, and in the examples below we take $B = B'$.

\section{Results}\label{sec:results}
\subsection{Reflections from the wall in 2D}
\begin{figure*}[tp]
\centering
\includegraphics[width = 1.0\textwidth]{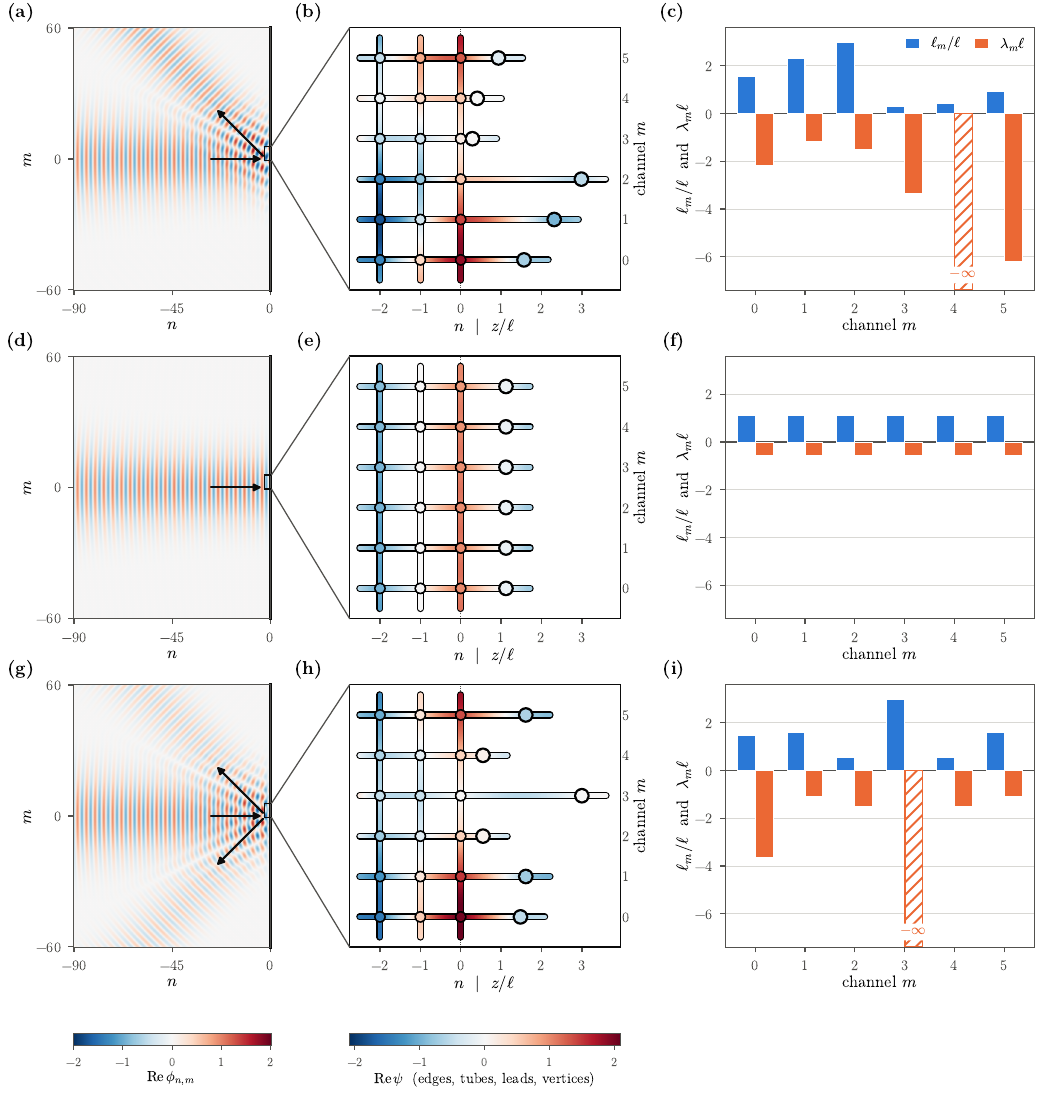}
\caption{Wall designs for a Gaussian beam at normal incidence ($k\ell = \pi/3$, wall period $L = 6\ell$, beam width $\sigma = 0.09/\ell$): (a)--(c) reflection at $+45^\circ$, (d)--(f) complete absorption and (g)--(i) splitting into $\pm45^\circ$. (a), (d), (g) Total field $\mathrm{Re}\,\phi_{n,m}$; arrows mark the incident and reflected directions. (b), (e), (h) One period of the wall, boxed in (a), (d), (g), with the field on the lattice edges and vertices, in the tubes and on the truncated leads; large circles mark the channel vertices. (c), (f), (i) Tube lengths $\ell_m/\ell$ (blue) and vertex parameters $\lambda_m\ell$ (orange) from Eqs.~\eqref{eq:lambda}--\eqref{eq:Gamma}; hatched bars mark Dirichlet channels, $\lambda_m = -\infty$. Lengths are defined modulo $\pi/k$. Colour scales are linear.}\label{fig:results}
\end{figure*}

\begin{figure*}[tp]
\centering
\includegraphics[width = 1.0\textwidth]{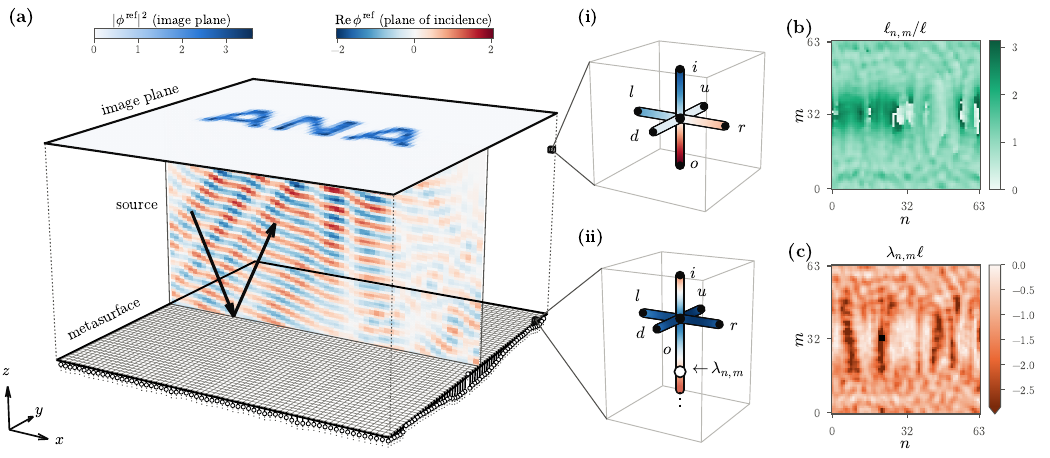}
\caption{Image formation by a two-dimensional metasurface terminating a cubic lattice at $k\ell = 1$, with a period of $64\ell\times64\ell$ ($4096$ channels). (a) A plane wave incident in the $x$-$z$ plane is reflected to form the letters ``ANA'' as thanks to our research funding the Adelaide-Nottingham Alliance; ($|U_{n,m}| = 1$ on the letters and $0$ elsewhere. The reflected field $\mathrm{Re}\,\phi^{\refl}$ is shown on the plane of incidence and $|\phi^{\refl}|^2$ on one period of the image plane; the tubes and truncated leads face down along two edges of the metasurface. (i) A lattice vertex and (ii) a vertex of the metasurface, marked by the small unit cells on the right-hand edge of the lattice, with their edges $l,r,d,u,i,o$ and the total field $\mathrm{Re}\,\psi$ plotted inside the edges; in (ii) the edge $o$ is the tube, which ends in the channel vertex $\lambda_{n,m}$. (b) Tube lengths $\ell_{n,m}/\ell$ and (c) vertex parameters $\lambda_{n,m}\ell$ (Appendix~\ref{app:3d}); lengths are defined modulo $\pi/k$. Colour scales are linear; (c) saturates below $\lambda\ell = -2.85$, and the black square marks the single Dirichlet channel.}\label{fig: 3D contours}
\end{figure*}

See Fig.~\ref{fig:results}, where we consider a non-trivial reflection angle, complete absorption and beam splitting for an incident beam at normal incidence. The beams are superpositions of Bloch waves with a Gaussian weight in the tangential wavenumber~\cite{lawrie2022}; the wall is designed for the central component and held fixed, and every other component is reflected according to the scattering problem of that wall (Appendix~\ref{app:beams}).

\subsection{Wave scattering in 3D and image forming}\label{sec:3D}

The same construction applies to a cubic lattice terminated by a two-dimensional metasurface. The vertices $(n,m,p)$ sit at $x = n\ell$, $y = m\ell$ and $z = p\ell$, the lattice occupies $p\ge0$, and the wall lies in the plane $p = 0$. Each vertex has six edges, $e\in\{l,r,d,u,i,o\}$, pointing along $\mp x$, $\mp y$ and $\pm z$ [Fig.~\ref{fig: 3D contours}(a)(i)], and each wall vertex $(n,m,0)$ is connected through its edge $o$ to a tube of length $\ell_{n,m}$ ending in a vertex with parameter $\lambda_{n,m}$ [Fig.~\ref{fig: 3D contours}(a)(ii)]. The eigenvalue problem~\eqref{eq:eig} now gives $\cos(\kappa_x\ell) + \cos(\kappa_y\ell) + \cos(\kappa_z\ell) = 3\cos(k\ell)$, a wall with a period of $N\times M$ channels scatters an incident wave into $N\times M$ orders, and $\Gamma_{n,m}$ is again given by Eq.~\eqref{eq:Gamma}, now with two-dimensional Fourier sums (Appendix~\ref{app:3d}).

To form an image in the plane $p = p_f$, we take the desired image $U_{n,m}$ on one period, compute its discrete Fourier coefficients $\tilde U_{\eta,\mu}$, and propagate each component back from the image plane to the wall, to give
\begin{equation}\label{eq:image}
B^{(\eta,\mu)} = B\,\tilde U_{\eta,\mu}\,\ee^{-i\kappa_z^{(\eta,\mu)}p_f\ell}\,H_{\eta,\mu},
\end{equation}
where $H_{\eta,\mu} = 1$ for propagating orders and $H_{\eta,\mu} = 0$ for evanescent ones. Figure~\ref{fig: 3D contours} shows the result for the letters ``ANA'' (for the Adelaide--Nottingham Alliance), together with the tube lengths and vertex parameters of every channel.

\section{Discussion and conclusions}\label{sec:conclusion}

We have shown that the boundary of a quantum graph lattice can be designed exactly to produce a prescribed reflected field, without optimisation. The required boundary condition at each channel reduces to a single complex ratio, $\Gamma_m$, of returning to outgoing amplitudes, Eq.~\eqref{eq:Gamma}, which follows from the incident and desired reflected fields through two discrete Fourier sums. This ratio is inverted in closed form for a tube length and a vertex parameter, Eqs.~\eqref{eq:lambda} and \eqref{eq:length}. The wall is passive whenever $|\Gamma_m|\le1$, that is, whenever every channel absorbs rather than emits energy. We demonstrated the construction for a one-dimensional metasurface terminating a two-dimensional lattice, which steers, splits or perfectly absorbs Gaussian beams, and for a two-dimensional metasurface terminating a three-dimensional lattice, which forms a prescribed image.

The model maps directly onto physical waveguide networks. Quantum graphs describe thin acoustic, elastic and electromagnetic waveguides~\cite{Kuchment2002} and can be built as acoustic pipe networks~\cite{lawrie2024} and as microwave networks of coaxial cables~\cite{Hul2004,lawrie2026}. The tube lengths can be folded into a compact wall by coiling~\cite{Liang2012}, and walls of acoustic cavities of this kind have already been built and tested~\cite{tian2022motor}. The vertex parameter, $\lambda_m$, corresponds to a local mass or spring, or more generally to any lumped scatterer, and the semi-infinite leads correspond to anechoic terminations or absorbing material. In such realisations, $\lambda_m$ usually depends on frequency (for a mass, $\lambda\propto-\omega^2$), which does not affect the design at the operating frequency. Because a single real parameter per channel controls loss and a single length controls phase, the wall is naturally reconfigurable. Tuning $\lambda_m$, for instance with adjustable masses, springs or resonators, switches between absorbing, steering and splitting states while the structure remains passive throughout. In contrast to the binary and phase-quantised surfaces discussed in Sec.~\ref{sec:intro}, the required settings are obtained in closed form rather than by a combinatorial search.

Several extensions follow directly. Other lattices and vertex conditions only change the Bloch amplitudes of Appendix~\ref{app:bloch}. Adding further tubes or vertices to each channel provides more parameters per channel, which can be used to prescribe richer fields, for several frequencies or several incident waves at once. The same reasoning applies to transmission through a structured layer~\cite{lawrie2023interface,Lawrie2026transmission}, and the semi-infinite leads could be replaced by a second lattice into which the wall transmits.

\begin{acknowledgments}
Any data and code generated for this paper can be made available upon reasonable request.
This research was funded by the Universities of Adelaide and Nottingham through the Adelaide-Nottingham Alliance (ANA). Tristan Lawrie would like to acknowledge the financial support of the META4D program grant (EP/Y015673/1), as well as many useful discussion with Simon Horsley. 

\end{acknowledgments}

\appendix

\section{The channel condition and its inversion}\label{app:channel}

\paragraph*{Vertex scattering matrix.} On an edge $e$ leaving a vertex we write the solution of Eq.~\eqref{eq:helmholtz} as $\psi_e(z) = a_e^{\out}\ee^{ikz} + a_e^{\inc}\ee^{-ikz}$, with $z$ measured from the vertex. Continuity at the vertex gives $a_e^{\out} + a_e^{\inc} = \phi$ for every edge, where $\phi$ is the vertex value, and the condition $\sum_e\psi_e'(0) = \lambda\phi$ gives $ik\sum_e(a_e^{\out} - a_e^{\inc}) = \lambda\phi$. Eliminating $\sum_e a_e^{\out} = v\phi - \sum_e a_e^{\inc}$ gives
\begin{equation} \label{philambda}
\phi = \frac{2ik}{vik-\lambda}\sum_e a_e^{\inc}, \qquad a_e^{\out} = \phi - a_e^{\inc},
\end{equation}
which is Eq.~\eqref{Scattering Definition}; see also Ref.~\cite{lawriethesis}. For the channel vertices, $v = 2$, and the scattering matrix has the form
\begin{equation}\label{eq:rhotau}
\boldsymbol{S}_m = \begin{pmatrix} \rho_m & \tau_m \\ \tau_m & \rho_m \end{pmatrix}, \quad
\rho_{m} = \frac{\lambda_m}{2ik - \lambda_m}, \quad \tau_m = \frac{2ik}{2ik - \lambda_m},
\end{equation}
with $|\rho_m|^2 + |\tau_m|^2 = 1$.

\paragraph*{Channel condition.} Let $\beta^{\inc}_{0,m,l}$ and $\beta^{\inc}_{0,m,r}$ be the amplitudes arriving at the channel vertex from the tube and from the lead, and $\beta^{\out}_{0,m,l}$, $\beta^{\out}_{0,m,r}$ those leaving it [Fig.~\ref{fig: Reflection Schematic}(b)], so that $(\beta^{\out}_{0,m,l},\beta^{\out}_{0,m,r})^T = \boldsymbol{S}_m(\beta^{\inc}_{0,m,l},\beta^{\inc}_{0,m,r})^T$. Propagation along the tube gives $\beta^{\inc}_{0,m,l} = \ee^{ik\ell_m}\alpha^{\out}_{0,m,r}$ and $\alpha^{\inc}_{0,m,r} = \ee^{ik\ell_m}\beta^{\out}_{0,m,l}$, so that
\begin{equation}\label{eq:lead}
\alpha^{\inc}_{0,m,r} = \ee^{2ik\ell_m}\rho_m\,\alpha^{\out}_{0,m,r} + \ee^{ik\ell_m}\tau_m\,\beta^{\inc}_{0,m,r}.
\end{equation}
A nonzero $\beta^{\inc}_{0,m,r}$ would inject energy into the lattice from the lead. For a passive wall $\beta^{\inc}_{0,m,r} = 0$, and Eq.~\eqref{eq:lead} reduces to Eq.~\eqref{Governing equation}. The wave leaving through the lead, $\beta^{\out}_{0,m,r} = \ee^{ik\ell_m}\tau_m\alpha^{\out}_{0,m,r}$, carries the energy flux $k|\tau_m|^2|\alpha^{\out}_{0,m,r}|^2 = k(|\alpha^{\out}_{0,m,r}|^2 - |\alpha^{\inc}_{0,m,r}|^2)$ out of the lattice.

\paragraph*{Inversion.} Since $k$, $\ell_m$ and $\lambda_m$ are real, the modulus of Eq.~\eqref{Governing equation} gives $|\rho_m|^2 = \lambda_m^2/(\lambda_m^2+4k^2) = |\Gamma_m|^2$, which is solved by Eq.~\eqref{eq:lambda}. The phase of Eq.~\eqref{Governing equation} gives $\ee^{2ik\ell_m} = \Gamma_m/\rho_m = \Gamma_m(2ik/\lambda_m - 1)$. From Eq.~\eqref{eq:rhotau}, $\cos(\arg\rho_m) = -|\rho_m|$ and $\sgn(\arg\rho_m) = -\sigma_m$, so that $\arg\rho_m = -\sigma_m\arccos(-|\Gamma_m|)$, which gives Eq.~\eqref{eq:length}. Equivalently, $\ell_m = (2ik)^{-1}\ln[\Gamma_m(2ik/\lambda_m-1)] + q\pi/k$, where the argument of the logarithm has unit modulus, so that $\ell_m$ is real.

\section{Bloch waves of the square lattice}\label{app:bloch}

With the amplitudes around a vertex ordered as $(l,r,d,u)$ and Bloch conditions $\phi_{n+1,m} = \ee^{i\kappa_x\ell}\phi_{n,m}$ and $\phi_{n,m+1} = \ee^{i\kappa_y\ell}\phi_{n,m}$ on the vertex values, the phase matrix in Eq.~\eqref{eq:eig} is $\boldsymbol{P} = \mathrm{diag}\{\boldsymbol{p}_x,\boldsymbol{p}_y\}$, with
\begin{equation}\label{eq:P}
\boldsymbol{p}_j = \ee^{ik\ell}
\begin{pmatrix}
0 & \ee^{-i\kappa_j \ell} \\
\ee^{i\kappa_j \ell} & 0
\end{pmatrix}.
\end{equation}
The dispersion relation~\eqref{eq:dispersion2D} gives
\begin{equation}\label{eq:kx}
\kappa_x(k,\kappa_y) = \frac{1}{\ell}\arccos\big(2\cos(k\ell) - \cos(\kappa_y\ell)\big),
\end{equation}
and we take the branch with $\mathrm{Im}\,\kappa_x\ge 0$ and, for propagating waves with $0<k\ell<\pi$, $0<\kappa_x\ell<\pi$. Then $\ee^{+i\kappa_x n\ell}$ either carries energy towards the wall or decays in that direction, and $\ee^{-i\kappa_x n\ell}$ either carries energy away from the wall or decays into the lattice.

Continuity at the vertex as specified in Appendix \ref{app:channel} requires $a^{\out}_e + a^{\inc}_e = \phi$ with $\phi$ given in (\ref{philambda}). Equation~\eqref{eq:eig} is homogeneous and fixes $\mathbf{a}^{\inc}$ only up to a constant factor, which we choose such that $\phi = 1$. 
With this normalisation, we have $\psi_e(\ell) = a^{\out}_e\ee^{ik\ell} + a^{\inc}_e\ee^{-ik\ell} = \ee^{i\theta_e} \psi(0) = \ee^{i\theta_e}$, which yields
\begin{equation}\label{eq:amps}
a_e^{\out} = \frac{\ee^{i\theta_e} - \ee^{-ik\ell}}{2i\sin k\ell}, \qquad
a_e^{\inc} = \frac{\ee^{ik\ell} - \ee^{i\theta_e}}{2i\sin k\ell},
\end{equation}
where $\theta_e = (-\kappa_x\ell,\kappa_x\ell,-\kappa_y\ell,\kappa_y\ell)$ for $e = (l,r,d,u)$ is the Bloch phase of the neighbouring vertex. These amplitudes satisfy $a^{\out}_e + a^{\inc}_e = 1$ for every $e$, and the vertex condition $\sum_e(a^{\out}_e - a^{\inc}_e) = 0$ reduces to $\sum_e\ee^{i\theta_e} = 4\cos k\ell$, which is the dispersion relation~\eqref{eq:dispersion2D}. The amplitudes $b_e$ of the waves travelling away from the wall follow with $\kappa_x\to-\kappa_x$. The field on an edge from a vertex position $(n,m)$ in direction $e$ is obtained as
\begin{equation}\label{eq:edgeprofile}
\psi_{n,m,e}(z) = \phi_{n,m} \frac{e^{i\theta_e}  \sin kz + \sin k(\ell-z)}{\sin k\ell},
\end{equation}
valid for $k\ell\notin\pi\mathbb{Z}$ with $\phi_{(n,m)}$, the amplitude at vertex $(n,m)$. A Bloch wave of vertex amplitude $A$ carries the energy flux $k(|a^{\out}_r|^2 - |a^{\inc}_r|^2)|A|^2 = k|A|^2\sin(\kappa_x\ell)/\sin(k\ell)$ along each horizontal edge, and evanescent waves carry none. Energy flows along the group velocity, $\nabla_{\boldsymbol\kappa}\omega \propto (\sin\kappa_x\ell,\sin\kappa_y\ell)$, normal to the iso-frequency contour, at the angle $\theta = \arctan(\sin\kappa_y\ell/\sin\kappa_x\ell)$ to the $x$ axis; because of the anisotropy of the lattice this generally differs from the direction of $(\kappa_x,\kappa_y)$.

\section{The design rule}\label{app:design}

The prescribed field at the vertices is the incident Bloch wave plus the reflected orders~\eqref{eq:orders},
\begin{equation}\label{eq:vertexfield}
\phi_{n,m} = A\,\ee^{i(\kappa_x^{(0)}n+\kappa_y^{(0)}m)\ell} + \sum_{j=0}^{M-1} B^{(j)}\ee^{i(-\kappa_x^{(j)}n+\kappa_y^{(j)}m)\ell},
\end{equation}
and the field on every edge follows from Eq.~\eqref{eq:edgeprofile}. On the tube attached to the wall vertex $(0,m)$, the outgoing and incoming amplitudes are the corresponding sums of the edge amplitudes $a_r$ and $b_r$ of Eq.~\eqref{eq:amps},
\begin{equation}\label{eq:alphas}
\alpha^{\out/\inc}_{0,m,r} = A\,a_r^{\out/\inc,(0)}\ee^{i\kappa_y^{(0)}m\ell} + \sum_{j}B^{(j)}b_r^{\out/\inc,(j)}\ee^{i\kappa_y^{(j)}m\ell},
\end{equation}
see Fig.~\ref{fig: Reflection Schematic}(b).
Inserting Eq.~\eqref{eq:amps} and using Eqs.~\eqref{eq:Phi} and \eqref{eq:Phitilde}, these become $\alpha^{\out}_{0,m,r} = (\tilde\Phi_m - \ee^{-ik\ell}\Phi_m)/(2i\sin k\ell)$ and $\alpha^{\inc}_{0,m,r} = (\ee^{ik\ell}\Phi_m - \tilde\Phi_m)/(2i\sin k\ell)$, whose ratio is Eq.~\eqref{eq:Gamma}.

The energy flux absorbed by channel $m$ is
\begin{equation}\label{eq:Pm}
P_m = k\big(|\alpha^{\out}_{0,m,r}|^2 - |\alpha^{\inc}_{0,m,r}|^2\big) = \frac{k}{\sin k\ell}\,\mathrm{Im}\big(\Phi_m^*\tilde\Phi_m\big),
\end{equation}
so the wall is passive, $|\Gamma_m|\le1$, if and only if $P_m\ge0$ for every $m$. When the reflected field is prescribed up to an overall amplitude, $B^{(j)} = B\hat B^{(j)}$, $\max_m|\Gamma_m|$ grows with $B$ from its value for the absorbing wall at $B = 0$, and reaches unity at the critical amplitude $B'$.

The prescribed field satisfies Eq.~\eqref{eq:helmholtz} on every edge and the vertex condition at every lattice vertex, and once Eq.~\eqref{Governing equation} holds it also satisfies the condition at every channel vertex; being unique, it is the response of the designed wall. Conversely, for a given wall $(\ell_m,\lambda_m)$ and incident amplitude $A$, imposing Eq.~\eqref{Governing equation} at all $M$ channels gives a linear system for the $M$ amplitudes $B^{(j)}$. This forward problem returns the prescribed amplitudes for every designed wall and is used to construct the beams of Appendix~\ref{app:beams}.

\section{Gaussian beams and the perfect absorber}\label{app:beams}

Following Ref.~\cite{lawrie2022}, a beam is constructed by integrating the Bloch solutions over the tangential wavenumber with a Gaussian weight centred on $\kappa_y'$,
\begin{equation}\label{eq:beam}
\phi^{\inc}_{n,m} = \int \dd q\, g(q)\, \ee^{i[\kappa_x(q)n + qm]\ell}, \qquad g(q) = \ee^{-(q-\kappa_y')^2/2\sigma^2},
\end{equation}
with $\kappa_x(q)$ from Eq.~\eqref{eq:kx}. The wall is designed for the central component $q = \kappa_y'$ and held fixed. Every other component is scattered into the orders $q - 2\pi j/L$, with amplitudes $B^{(j)}(q)$ given by the forward problem of Appendix~\ref{app:design}, and the reflected beam is
\begin{equation}\label{eq:beamref}
\phi^{\refl}_{n,m} = \int \dd q\, g(q) \sum_{j=0}^{M-1} B^{(j)}(q)\, \ee^{i[-\kappa_x^{(j)}(q)\,n\ell + (q - 2\pi j/L)\,m\ell]}.
\end{equation}
Incident and reflected beams together form an exact solution of the graph with the designed wall. For $k\ell = \pi/3$ and $L = 6\ell$, the orders $j = 1$ and $j = 5$ of a normally incident wave have $\kappa_x^{(j)} = |\kappa_y^{(j)}| = k$, so that they leave the wall at exactly $\mp45^\circ$, while all other nonspecular orders are evanescent.

For perfect absorption, $B^{(j)} = 0$, we have $\tilde\Phi_m = \ee^{i\kappa_x^{(0)}\ell}\Phi_m$, and Eq.~\eqref{eq:Gamma} gives the same value for every channel,
\begin{equation}\label{eq:rhoprime}
\Gamma_m = \rho' = \ee^{ik\ell}\,\frac{\sin\big[(k-\kappa_x^{(0)})\ell/2\big]}{\sin\big[(k+\kappa_x^{(0)})\ell/2\big]}.
\end{equation}
The absorbing wall is therefore uniform, and it is always passive, since $|\rho'|<1$ is equivalent to $\sin k\ell\,\sin\kappa_x^{(0)}\ell>0$. For a wave incident with $\kappa_y^{(0)} = k$, Eq.~\eqref{eq:dispersion2D} gives $\kappa_x^{(0)} = k$, so that $\rho' = 0$ and $\lambda' = 0$: a wave travelling along the lattice diagonal is absorbed completely by open leads attached to the boundary vertices.

\section{The cubic lattice}\label{app:3d}

For the cubic lattice the amplitudes around a vertex are ordered as $(l,r,d,u,i,o)$, with $l,r$ along $\mp x$, $d,u$ along $\mp y$ and $i,o$ along $+z$ and $-z$. The Neumann--Kirchhoff scattering matrix has elements $S_{e,e'} = 1/3 - \delta_{e,e'}$, and the phase matrix is $\boldsymbol{P} = \mathrm{diag}\{\boldsymbol{p}_x,\boldsymbol{p}_y,\boldsymbol{p}_z\}$ with $\boldsymbol{p}_j$ as in Eq.~\eqref{eq:P}. The zeros of $\det[\mathbf{I}-\boldsymbol{P}\boldsymbol{S}]$ give the dispersion relation
\begin{equation}\label{3D dispersion equation}
\cos(\kappa_x\ell) + \cos(\kappa_y\ell) + \cos(\kappa_z\ell) = 3\cos(k\ell),
\end{equation}
and the eigenvectors are again given by Eq.~\eqref{eq:amps}, now with six Bloch phases $\theta_e$. A wall with a period of $N\times M$ channels scatters an incident wave with tangential wavenumbers $(\kappa_x^{(0)},\kappa_y^{(0)})$ into the orders $\kappa_x^{(\eta)} = \kappa_x^{(0)} - 2\pi\eta/(N\ell)$ and $\kappa_y^{(\mu)} = \kappa_y^{(0)} - 2\pi\mu/(M\ell)$, with $\kappa_z^{(\eta,\mu)}$ from Eq.~\eqref{3D dispersion equation} and $\mathrm{Im}\,\kappa_z\ge0$. The prescribed field at the vertex $(n,m,p)$ is
\begin{equation}\label{eq:field3D}
\phi_{n,m,p} = \sum_{\eta,\mu} E^{(\eta,\mu)}_{n,m}\Big[A^{(\eta,\mu)}\ee^{-i\kappa_z^{(\eta,\mu)}p\ell} + B^{(\eta,\mu)}\ee^{i\kappa_z^{(\eta,\mu)}p\ell}\Big],
\end{equation}
with $E^{(\eta,\mu)}_{n,m} = \ee^{i(\kappa_x^{(\eta)}n + \kappa_y^{(\mu)}m)\ell}$ and sums over $0\le\eta<N$, $0\le\mu<M$. The incident field may itself contain several orders, and the field on every edge follows from Eq.~\eqref{eq:edgeprofile}. The tube of channel $(n,m)$ is the edge $o$ of the wall vertex $(n,m,0)$. Repeating the steps of Appendix~\ref{app:design}, $\Gamma_{n,m}$ is given by Eq.~\eqref{eq:Gamma} with the two-dimensional Fourier sums
\begin{equation}\label{eq:Gamma3D}
\begin{split}
\Phi_{n,m} &= \sum_{\eta,\mu}E^{(\eta,\mu)}_{n,m}\big[A^{(\eta,\mu)} + B^{(\eta,\mu)}\big],\\
\tilde\Phi_{n,m} &= \sum_{\eta,\mu}E^{(\eta,\mu)}_{n,m}\big[A^{(\eta,\mu)}\ee^{i\kappa_z^{(\eta,\mu)}\ell} + B^{(\eta,\mu)}\ee^{-i\kappa_z^{(\eta,\mu)}\ell}\big],
\end{split}
\end{equation}
the field at the wall vertex and its continuation to the missing layer $p = -1$. Equations~\eqref{eq:lambda} and \eqref{eq:length} then give $\lambda_{n,m}$ and $\ell_{n,m}$.

For an incident plane wave and the choice~\eqref{eq:image}, with $U_{n,m} = \sum_{\eta,\mu}\tilde U_{\eta,\mu}\ee^{-2\pi i(\eta n/N + \mu m/M)}$, the reflected field in the image plane is $B\,\ee^{i(\kappa_x^{(0)}n + \kappa_y^{(0)}m)\ell}$ times the band-limited version of $U_{n,m}$, in which only the Fourier components inside the iso-frequency surface~\eqref{3D dispersion equation} survive. Its intensity therefore reproduces $|U_{n,m}|^2$ for any angle of incidence. Because the wall and the incident wave are periodic, so is the image; Fig.~\ref{fig: 3D contours}(a) shows one period of the image plane, directly above one period of the wall.

\bibliographystyle{unsrt}
\bibliography{passive_wall_refs}

\end{document}